\documentclass[11pt,twoside,a4paper,cmspaper,final]{cms-tdr}
\def\svnVersion{264323MP}\def\cmsCernNo{2015-NNN}\def\cmsCernDate{\today}\def\cmsMessage{Submitted to JINST}

\begin{document}\cmsNoteHeader{}

\hyphenation{had-ron-i-za-tion}
\hyphenation{cal-or-i-me-ter}
\hyphenation{de-vices}

\RCS$Revision: 264263 $
\RCS$HeadURL: svn+ssh://svn.cern.ch/reps/tdr2/notes/tpoehlse_001/trunk/tpoehlse_001.tex $
\RCS$Id: tpoehlse_001.tex 264263 2014-10-17 11:08:02Z alverson $
\newlength\cmsFigWidth
\ifthenelse{\boolean{cms@external}}{\setlength\cmsFigWidth{0.85\columnwidth}}{\setlength\cmsFigWidth{0.4\textwidth}}
\ifthenelse{\boolean{cms@external}}{\providecommand{\cmsLeft}{top\xspace}}{\providecommand{\cmsLeft}{left\xspace}}
\ifthenelse{\boolean{cms@external}}{\providecommand{\cmsRight}{bottom\xspace}}{\providecommand{\cmsRight}{right\xspace}}
\renewcommand{\cmsCollabName}{Tracker group of the CMS Collaboration}
\cmsNoteHeader{tpoehlse\_001} 
\title{Trapping in irradiated p-on-n silicon sensors at fluences anticipated at the HL-LHC outer tracker}

\author*[hh]{Thomas Poehlsen}

\date{\today}

\abstract{
The degradation of signal in silicon sensors is studied under conditions expected at the CERN High-Luminosity LHC.
200~$\mu$m thick n-type silicon sensors are irradiated with protons of different energies to fluences of up to $3 \cdot 10^{15}$~neq/cm$^2$. Pulsed red laser light with a wavelength of 672~nm is used to generate electron-hole pairs in the sensors. The induced signals are used to determine the charge collection efficiencies separately for electrons and holes drifting through the sensor. The effective trapping rates are extracted by comparing the results to simulation.
The electric field is simulated using Synopsys device simulation assuming two effective defects. 
The generation and drift of charge carriers are simulated in an independent simulation based on PixelAV. 
The effective trapping rates are determined from the measured charge collection efficiencies and the simulated and measured time-resolved current pulses are compared.
The effective trapping rates determined for both electrons and holes are about 50\% smaller than those obtained using standard extrapolations of studies at low fluences and suggests an improved tracker performance over initial expectations.
}

\hypersetup{%
pdfauthor={T. Poehlsen},%
pdftitle={Trapping in irradiated p-on-n silicon sensors at fluences relevant for the HL-LHC outer tracker volumes},%
pdfsubject={CMS},%
pdfkeywords={silicon sensors, charge losses, trapping, charge collection efficiency}}

\maketitle 

\section{Introduction}

After the upgrade of the Large Hadron Collider (LHC) to the High-Luminosity LHC (HL-LHC), which is foreseen in 2022, the radiation damage the tracking detectors will experience increases significantly. For both the development of sensors with performance optimised for HL-LHC fluences and the development of Monte Carlo simulation, a quantitative description of signal loss in irradiated silicon sensors is needed, especially in the inner layers of the general-purpose experiments ATLAS~\cite{ATLAS} and CMS~\cite{CMS}. 

Radiation damage during operation will degrade tracker performance because of the generation of electrically active defects in the bulk of the silicon sensors \cite{Moll}. The main consequences are:
\begin{itemize} 
\item higher sensor leakage current leading to increased noise, heat generation, and power consumption;
\item a change in the space charge distribution reducing the active part of the sensor volume and requiring higher operating voltages;
\item trapping of charge carriers leading to lower signals and hence the degradation of the spatial resolution and the efficiency.
\end{itemize}
In this work we will concentrate on the effects of charge loss due to trapping. 

In previous work~\cite{Kramberger:2002, Kramberger:2002-2} charge loss was studied at 1 MeV neutron equivalent fluences\footnote{Neutron equivalent scaling is motivated by the leakage current, which was shown to be proportional to the non-ionising energy loss (NIEL)~\cite{Moll}. However, charge losses do not scale to the NIEL~\cite{Kramberger:2002, Kramberger:2002-2}.}, $\phi_{neq}$, of up to $2.4\cdot10^{14}$~neq/cm$^2$. This fluence range is relevant for large parts of the current CMS Tracker. The assumption of voltage-independent trapping rates was made. 
The measured signal currents are corrected with an exponential,

\begin{equation}
I_{corrected}(t)=I_{measured}(t) \cdot exp(t/\tau_{tr}),
\end{equation} 
with a free parameter $\tau_{tr}$ that is tuned so that the integrals of the corrected currents give equal charges for voltages above the full-depletion voltage. This method does not require information about the charge collection efficiency, as it could not be determined experimentally using the measurements that were taken. A linear dependence of the trapping rate on the fluence was found:
\begin{equation}
1/\tau_{e,h} = \beta_{e,h}(T) \cdot \phi_{neq} ,
\label{eq:kramberger}
\end{equation} 
where $1/\tau_{e,h}$ is the effective trapping rate and $\beta_{e,h}(T)$ is the temperature-dependent damage parameter for electrons and holes, respectively. For electrons (holes) at a sensor temperature of $-20\,^\circ$C a value of $\beta_e = (5.8 \pm 0.2)\cdot10^{-16}$cm$^2$/ns ($\beta_h = (8.2 \pm 0.2)\cdot10^{-16}$cm$^2$/ns) was found for sensors after charged-hadron irradiation~\cite{Kramberger:2002}. The quoted uncertainties do not include the 10~\% uncertainty associated with the dosimetry.
In studies at higher fluences \cite{Lan09,Poe10,Eber:2013} charge collection measurements were found to be in tension with those presented in Ref.~\cite{Kramberger:2002}, when considering only the voltage range where no charge multiplication is expected. It is therefore important to also determine effective trapping rates at the higher fluences expected at the HL-LHC of between 3$\cdot$$10^{14}$ and 3$\cdot$$10^{15}$~neq/cm$^2$ separately for electrons and for holes. 

This fluence range is expected after the collection of $3000$~fb$^{-1}$ of HL-LHC data at a radius in the range between $10$~cm and $60$~cm from the interaction point. Most of the HL-LHC fluence arises from pions created in pp collisions, with mechanisms for causing damage that are similar to those of protons. We used a simple trapping model that does not depend on local variation of the electric field, or on the charge carrier concentration. This is equivalent to an effective trapping rate that does not depend on the position in the sensor.

In the study presented here electron-hole pairs ($eh$-pairs) are generated using pulsed laser light of 672~nm wavelength on both the p$^+$ (front) and the n$^+$ (rear) side of pad sensors of p-on-n float-zone silicon.
Using this set-up, the charge collection efficiencies are determined, and effective trapping rates are extracted through simulation. The simulation is based on the expected electric field distributions in the presence of two defect levels. This method of describing the electric field is also used in Refs. \cite{Eber:2013, EVL1, EVL2, PixelAV}.
Finally, the extracted trapping rates are checked using pulsed laser light of 1062~nm wavelength to generate the $eh$-pairs.

\section{Sensors and measurement technique}
The p-on-n silicon pad sensors are produced by Hamamatsu Photonics\footnote{Hamamatsu webpage: \url{http://www.hamamatsu.com/}} from $\langle 1 0 0 \rangle$-oriented float-zone wafers of 200~$\mu$m thickness with an oxygen concentration of about $8$$\cdot$$10^{16}$~cm$^{-3}$. This oxygen concentration is similar to that of the oxygen-enriched float zone sensors studied in Ref.~\cite{PixelAV} (about $10^{17}$~cm$^{-3}$). Oxygen enriched sensors were also studied previously in Ref.~\cite{Kramberger:2002} where no significant dependence on the level of oxygen-enrichment was detected. The pad area of the sensors under study is $0.25$~cm$^2$. The full-depletion voltage before irradiation is about 90~V. Other measurements made using sensors from the same production run were reported in Refs.~\cite{Dierlamm:2012,Steinbrueck:2012, PoehlsenIEEE:2013, Poehlsen:2013, Erfle:2013, Scharf:2014}.

Electron-hole pairs are generated at either the front or the rear side of the p-on-n sensors. Pulsed laser light with a wavelength of 672~nm is used, which has a penetration depth in silicon of about 3.5~$\mu$m at the temperature used ($-20^\circ$C). The time-resolved charge collection measurements are performed in 10~V steps from 0~V up to 1000~V and analysed in detail at 600~V. 
A voltage of 600~V is chosen because it represents the upper limit for the outer tracker bias voltages arising from the current power supplies and safety limits on cables. The light pulses have a duration of about 60~ps full width half maximum, the number of $eh$-pairs generated by each pulse is about $10^6$, and the laser repetition rate is set to 200~Hz.

\begin{figure}
	\centering
	\includegraphics[width=7.8cm]{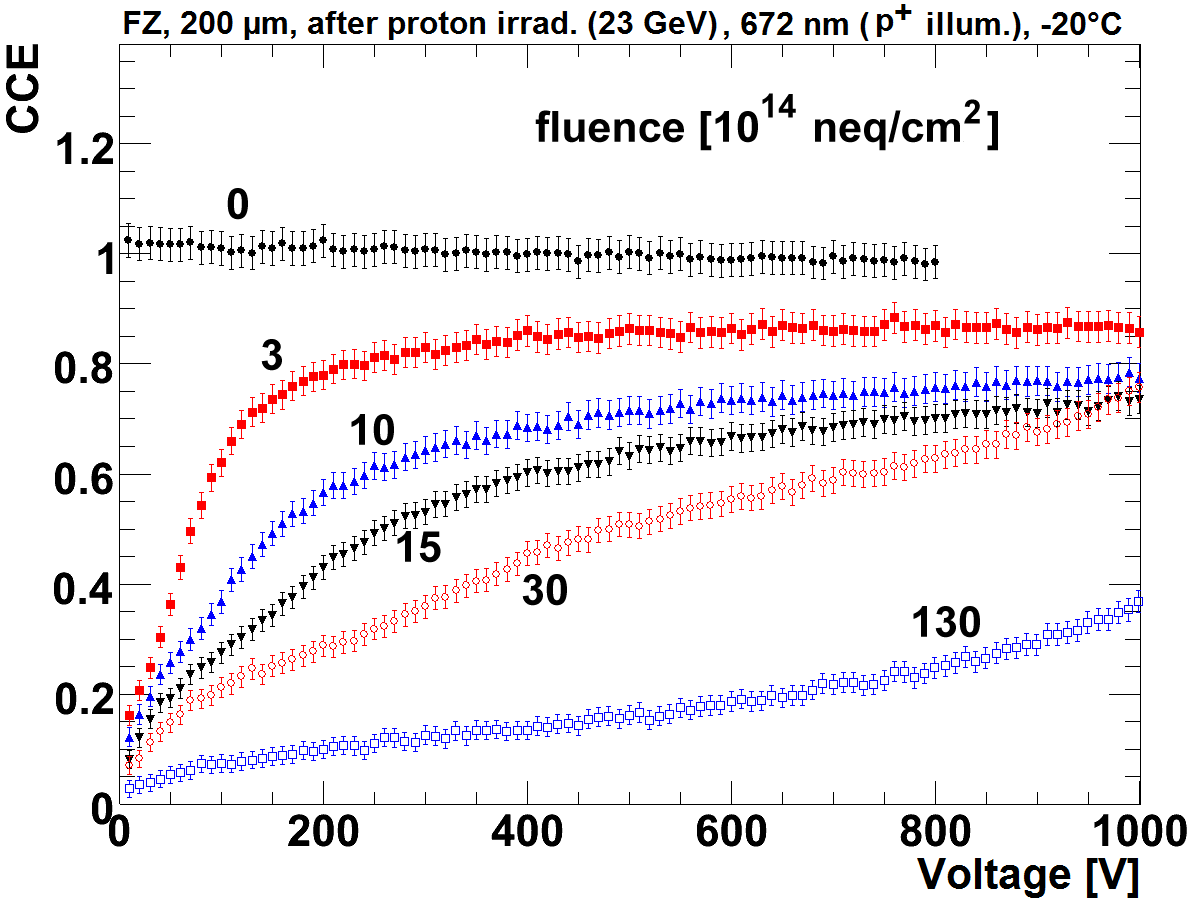}
	\includegraphics[width=7.8cm]{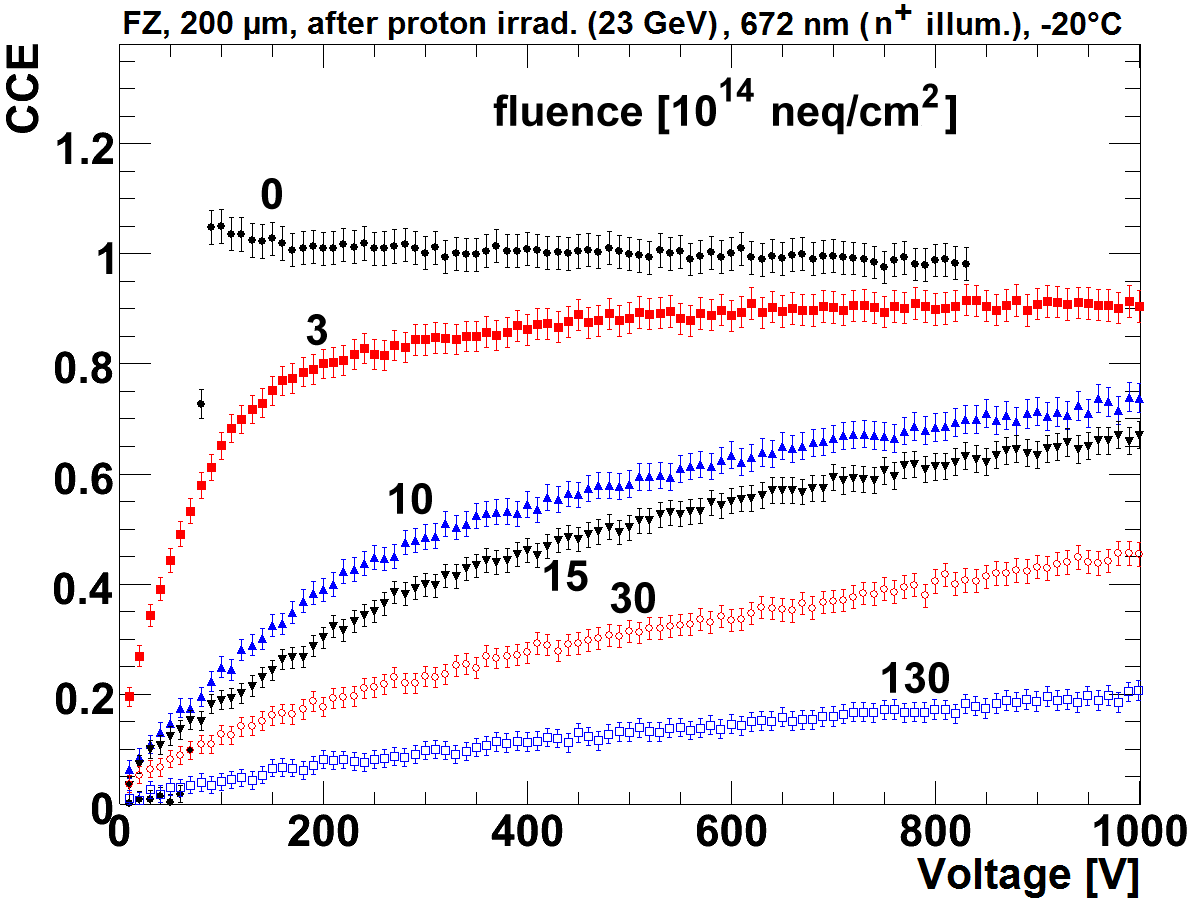}
  \caption{The CCE as a function of bias voltage is shown for 200~$\mu$m thick n-type sensors after different fluences of 23~GeV proton irradiation. Laser light of 672~nm wavelength is used to generate $eh$-pairs close to the p$^+$-side (left), so that the signals are dominated by electron drift, or close to the n$^+$-side (right), so that the signal is dominated by hole drift.}
	\label{fig:CCE_red_front}
\end{figure}

The current signal induced in the pad is read out by a digital oscilloscope with $1$~GHz bandwidth and 5~GHz sampling rate (Tektronix DPO 4104). The induced charge, $Q$, is calculated by integrating the time-resolved current signal over 30~ns, and the charge collection efficiency (CCE) is determined as the ratio of the collected charge after irradiation to that measured for a fully depleted non-irradiated reference sensor at 400~V bias (full-depletion voltage 90~V).
More details about the setup and the CCE determination can be found in Ref.~\cite{Poehlsen:2013}. The CCE as a function of bias voltage is shown in Figure~\ref{fig:CCE_red_front}. As expected for light with a short penetration depth the CCE is 0 for voltages below full depletion and 1 at voltages above full depletion, if the non-irradiated sensor is illuminated at the n$^+$ side. 

Five sensors were irradiated at the PS (CERN) with 23~GeV protons. These sensors were not cooled during irradiation, which took up to about two weeks for the highest fluence. They were investigated without additional annealing after irradiation. One sensor was irradiated at KIT (Karlsruhe) with 23~MeV protons. This sensor was cooled during irradiation to below 0\,$^\circ$C and investigated after 10~minutes of annealing at 60\,$^\circ$C. No significant dependence of the effective trapping times on annealing time has been observed in Ref.~\cite{Kramberger:2002}, such that no significant impact is expected as a result of the different annealing scenarios.

\section{Simulation of charge collection}

The electric field is calculated using Synopsys device simulation\footnote{Synopsys webpage: \url{http://www.synopsys.com}.}, assuming two effective traps: a deep acceptor, $A$, and a deep donor, $D$, with energy levels of $E_D=E_V + 0.48 $~eV and $E_A=E_C - 0.525$~eV~\cite{EVL2}, where $E_V$ and $E_C$ represent the energy levels of the valence band and the conduction band. 
Different defect concentrations and cross sections are used for the different irradiation types and for the different fluences 
The values relevant for the studies presented in this paper are reported in Table~\ref{tab:parameters}. Some values were extracted for silicon sensors after 24~GeV proton irradiation from grazing-angle test beam measurements that are described in Ref.~\cite{PixelAV}. In this work these values are used to describe measurements after 23~GeV proton irradiation.
The other values are taken from Ref.~\cite{Eber:2013}. They were tuned to describe capacitance, current, and time-resolved charge collection measurements on single-pad silicon sensors after 23~MeV proton irradiation. In this work they are used to describe measurements after 23~MeV proton irradiation. 

\begin{table}[b]
\centering
\begin{tabular}[b]{ccccccc}
\hline
$\phi_{neq}$ 	& $N_A$			& $N_D$			&  $\sigma_A^e$	&  $\sigma_D^e$	&  $\sigma_A^h$	&  $\sigma_D^h$	\\ 
 $[10^{14} $neq/cm$^2]$ 
							&	$[10^{14} $cm$^{-3}]$	& $[10^{14} $cm$^{-3}]$
										& $[10^{-15} $cm$^2]$		& $[10^{-15} $cm$^2]$		& $[10^{-15} $cm$^2]$		& $[10^{-15} $cm$^2]$		\\ \hline

$2$		(24 GeV) \cite{PixelAV}	& $6.8$		&	$10$			
											&	$6.6$ 	&	$6.6$	&	$1.65$ &	$6.6$ \\
$6$		(24 GeV) \cite{PixelAV}	& $16$		&	$40$			
											&	$6.6$ 	&	$6.6$ &	$1.65$ &	$1.65$ \\
$12$	(24 GeV) \cite{PixelAV,MS}	& $30$	&	$69$			
											&	$3.8$ 	&	$3.8$ &	$0.94$ &	$0.94$ \\ 

$3$		(23 MeV) \cite{Eber:2013}&  $4.2$	& $13$
											& $10$ 		&	$10$	& $10$	&	$10$\\
$10$	(23 MeV) \cite{Eber:2013}& $12.5$	&	$52$					
											& $10$ 		& $10$	&	$10$	&	$10$\\ \hline

 \end{tabular}
   \caption{The key parameter values used in the Synopsys device simulation. These include: donor and acceptor concentrations, $N_D$ and $N_A$, and their electron and hole capture cross sections, $\sigma^{e,h}_{D,A}$, for silicon sensors after irradiation with 24~GeV protons (top rows)~\cite{PixelAV}, and for sensors after irradiation with 23~MeV protons (bottom rows) \cite{Eber:2013}.}
   \label{tab:parameters}
\end{table}

\begin{figure}
	\centering
	\includegraphics[width=9cm]{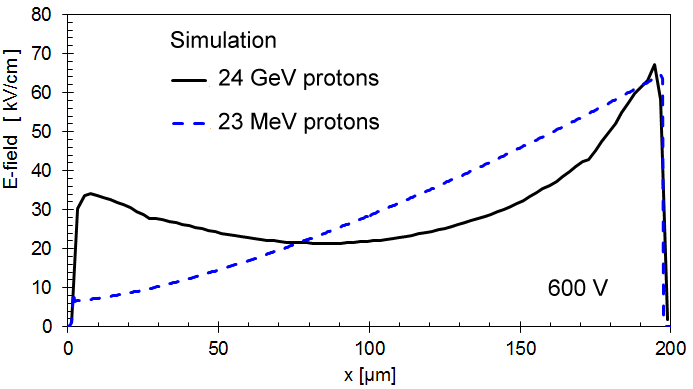}
  \caption{The simulated electric field at 600 V as a function of sensor depth, $x$, for a 200~$\mu$m thick n-type sensor after proton irradiation with different proton energies. The $p^+$ implant is at $x=0$~$\mu$m, and the $n^+$ implant is at $x=200$~$\mu$m. The field is calculated using parameters from Ref.~\cite{PixelAV} for irradiation with 24~GeV protons ($1.2\cdot10^{15}$~neq/cm$^2$) and Ref.~\cite{Eber:2013} for irradiation with 23~MeV protons ($1\cdot10^{15}$~neq/cm$^2$).
	}
	\label{fig:E-fields}
\end{figure}

In Figure~\ref{fig:E-fields} it can be seen that the electric field distribution is different for sensors after 23~MeV proton irradiation compared to that observed after 24~GeV proton irradiation.  The figure shows the electric field distribution for the defect values specified in Ref.~\cite{Eber:2013}, namely, for an irradiation with 23~MeV protons and a fluence of
 $10^{15}$~neq/cm$^2$, and for the defect values given in Ref.~\cite{PixelAV} for an irradiation with 24~GeV protons and a fluence of $1.2\cdot10^{15}$~neq/cm$^2$.

\begin{figure}
	\centering
	\includegraphics[width=7cm]{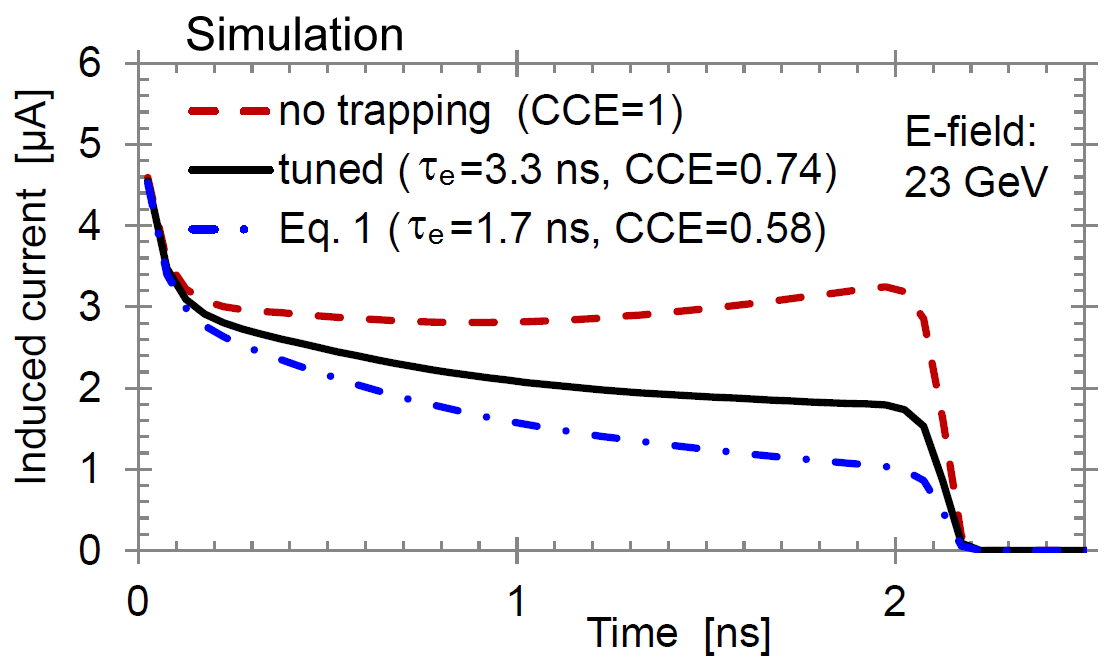}	
	\includegraphics[width=7cm]{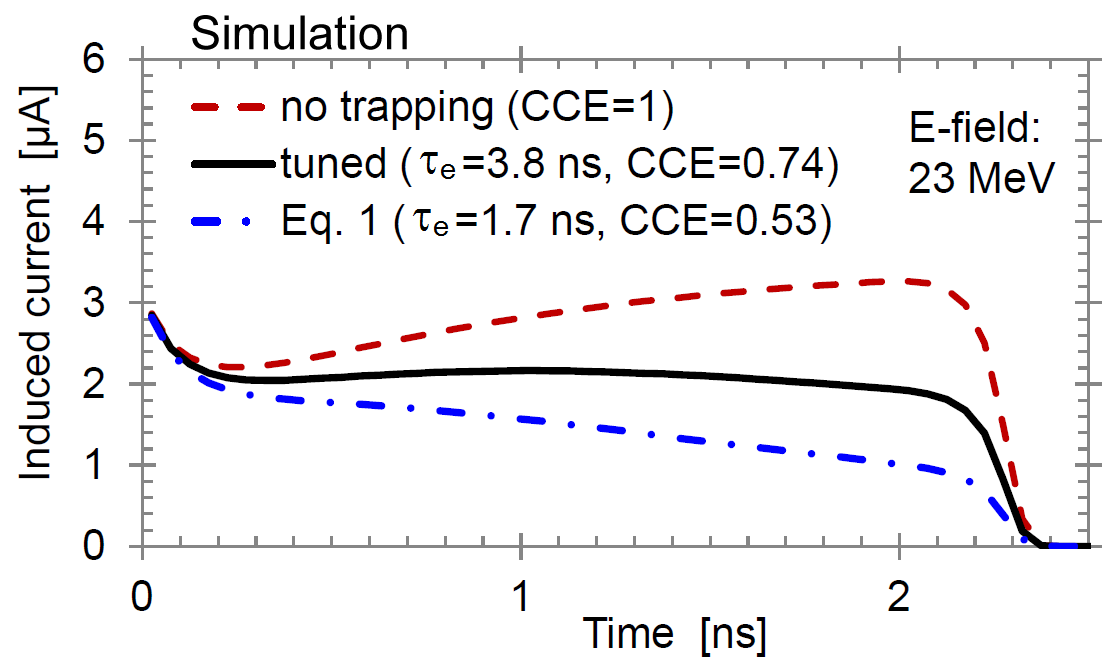}	
       \caption{Simulated current signals at 600~V bias for a proton-irradiated 200~$\mu$m thick n-type sensor after $40\,000$ $eh$-pairs are generated instantaneously close to the p$^+$-side. A penetration depth of 3.5~$\mu$m is used to simulate light with 672~nm wavelength. The signals are dominated by electron drift. Different electron trapping rates are used: no trapping, trapping tuned to $CCE=0.74$ (the value 0.74 is taken from Figure~\ref{fig:CCE_red_front} for $\phi_{neq}=10^{15}$~neq/cm$^2$), and trapping according to Equation~\ref{eq:kramberger} with $\phi_{neq}=10^{15}$~neq/cm$^2$. For the two proton energies the respective electric field distributions from Figure~\ref{fig:E-fields} are used. Left: simulation for 23~GeV protons. Right: simulation for 23~MeV protons.}
			\label{fig:TCT_1e15GeV} 
\end{figure}

PixelAV~\cite{PixelAV} is used to simulate the transport of charge carriers. The effective trapping rates in the simulation are assumed to be constant over the depth of the sensor.  Some modifications are made in order to describe the measurements reported here:
\begin{itemize}
\item Drift parameters are adjusted to describe the drift in $\langle 1 0 0 \rangle$-oriented silicon~\cite{Becker:2011};
\item Charges are generated at the front or the rear side of the sensor with a penetration depth of $3.5$~$\mu$m to simulate the charge generation by laser light of 672~nm wavelength. For simplicity the number of charges generated is fixed to $40\,000$ $eh$-pairs;
\item The induced signal is calculated using a linear weighting potential between front and rear contact;
\item The trapping rates are tuned iteratively to match the predicted CCE to the measured CCE.
\end{itemize}
In Figure~\ref{fig:TCT_1e15GeV} the resulting time-resolved current signals are shown for three different effective trapping rates and for two different electric fields. For the ``no trapping'' case, the integrated signals are $Q = 40\,000$ electrons, i.e. all charges are collected ($CCE = 1$). For the other cases the CCE decreases monotonically with increasing $1/\tau$. The effective trapping rate can be tuned to reproduce the measured CCE. 

\begin{figure}
	\centering
	\includegraphics[width=8cm]{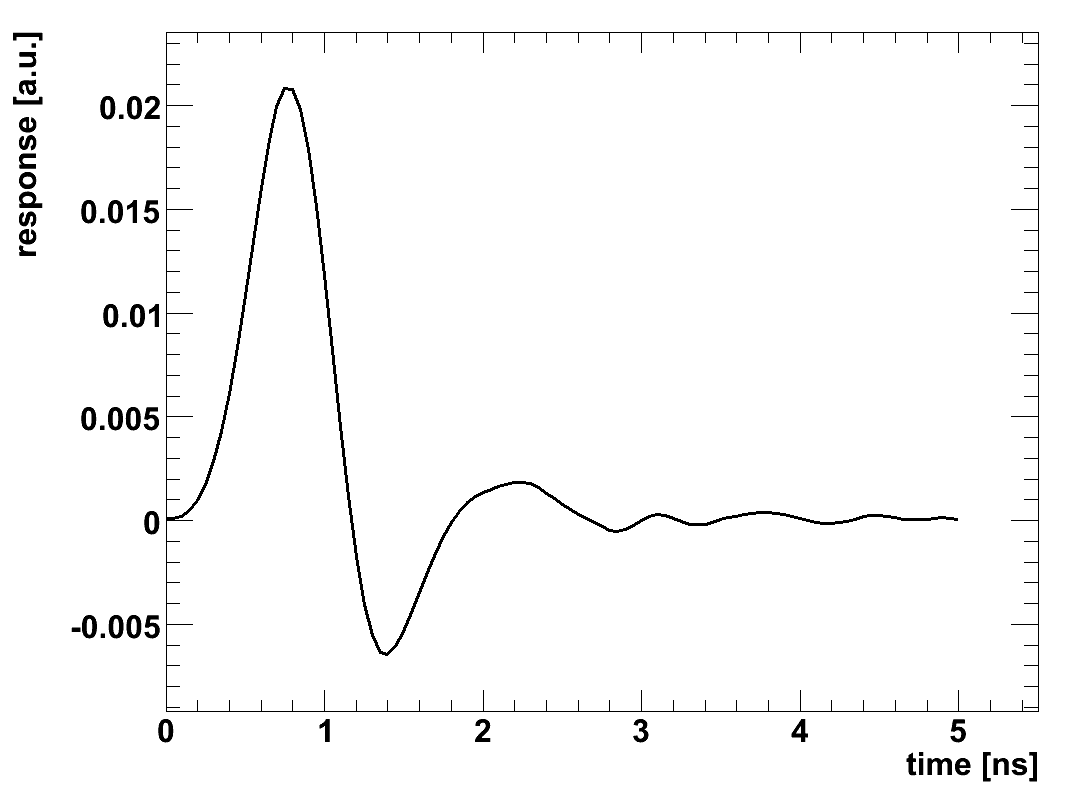}	
       \caption{Transfer characteristic of the circuit (response of the setup to a delta function for the sensor current).}
	\label{fig:response}
\end{figure}

\section{Comparison of measurements with simulation}

In order to compare the simulation to the sensor measurements, the electronic response of the experimental setup must be taken into account. This is achieved by convolving the simulated current signals with the response of the setup to a delta function; the latter is shown in Figure~\ref{fig:response}.
The response was extracted by studying the charge collection in non-irradiated sensors and has been reported in Ref.~\cite{Scharf:2014}.

\begin{figure}
	\centering
	\includegraphics[width=7.8cm]{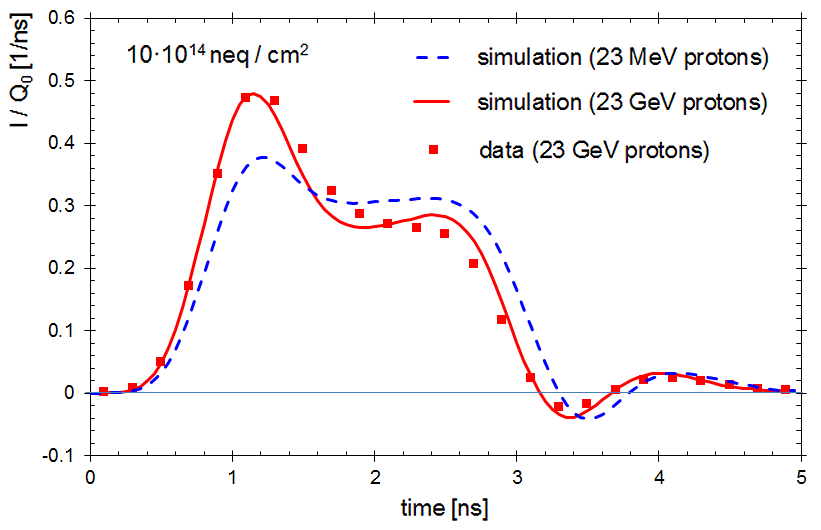}	
	\includegraphics[width=7.8cm]{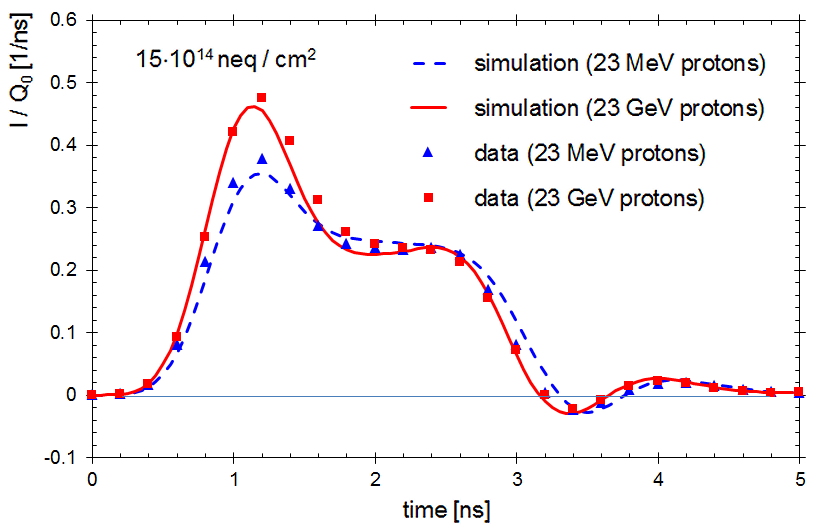}	
       \caption{Comparison of simulated and measured current signals, $I(t)$, at 600~V, normalised to the deposited charge,
			$Q_0$. The signals are mainly induced by electrons drifting from the front side to the rear side of the sensor.
			The results from a 200~$\mu$m thick n-type sensor are shown after 23~GeV proton irradiation to a fluence of $10^{15}$~neq/cm$^2$ (left) and $1.5\cdot10^{15}$~neq/cm$^2$ (right), and after 23~MeV proton irradiation to a fluence of $1.5\cdot10^{15}$~neq/cm$^2$ (right). At $10^{15}$~neq/cm$^2$ no sensors that had been irradiated with 23~MeV protons are available.
			}
	\label{fig:TCTcomparison}
\end{figure}

The simulated current signals, after the electronic response of the setup is taken into account, are compared to measured signals in Figure~\ref{fig:TCTcomparison}. The measurements are performed after proton irradiation of $10^{15}$~neq/cm$^2$ (23 GeV protons) and $1.5\cdot10^{15}$~neq/cm$^2$ (23 GeV protons and 23 MeV protons). A fluence of $1.5\cdot10^{15}$~neq/cm$^2$ is expected after the collection of $3000$~fb$^{-1}$ of HL-LHC data at a radius of $20$~cm from the interaction point.
In the simulation the effective trapping rates are adjusted so that the simulated CCE agrees with the measured CCE.
As seen in Figure~\ref{fig:E-fields} the expected electric field distribution is different for sensors after 23~MeV proton irradiation compared to that expected after 24~GeV proton irradiation. The same electric field is used for both fluences ($10^{15}$~neq/cm$^2$ and $1.5\cdot10^{15}$~neq/cm$^2$) as there are only limited data available on the two effective traps. For 23~GeV protons the field is based on parameters tuned to sensors irradiated to $1.2\cdot10^{15}$~neq/cm$^2$ using 24~GeV protons. For 23~MeV protons the field is based on parameters for the fluence $10^{15}$~neq/cm$^2$~\cite{Eber:2013} and 23~MeV protons. In Ref.~\cite{Eber:2013} the field has only been studied up to this fluence.

The simulated current signals are in good agreement with the measured currents, especially in light of the crude assumptions used in the simulation.
Taking into account the fact that the measured pulse shapes are quite different after 23~GeV proton irradiation compared to after 23~MeV proton irradiation (Figure~\ref{fig:TCTcomparison}), it is clear that different electric field distributions must be used for sensors that have undergone 23~MeV proton irradiation versus those that have undergone 23~GeV proton irradiation. However, even if quite different electric field distributions are used, the measured CCE is reproduced using similar trapping rates for the two cases (Figure~\ref{fig:TCT_1e15GeV}).

\section{Extracted trapping rates}
\begin{figure}
	\centering
	\includegraphics[width=7cm]{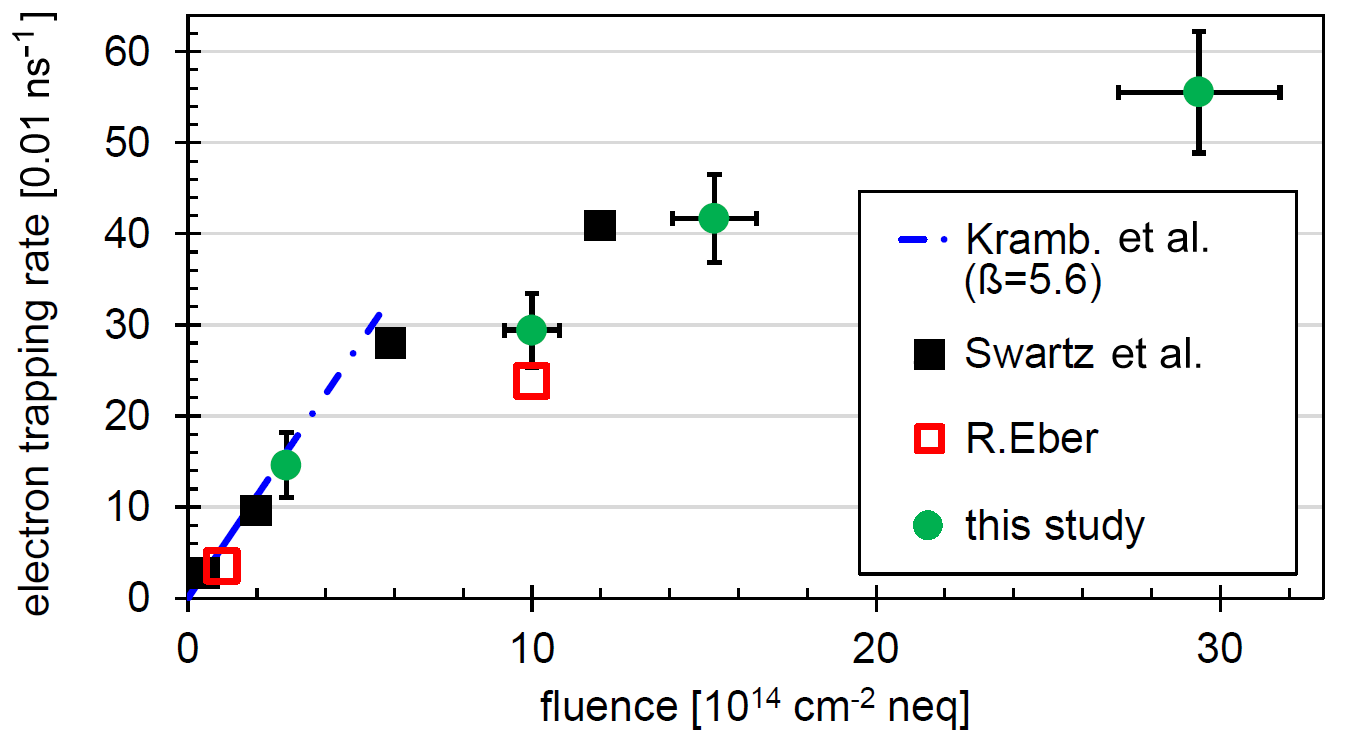}	
	\includegraphics[width=7cm]{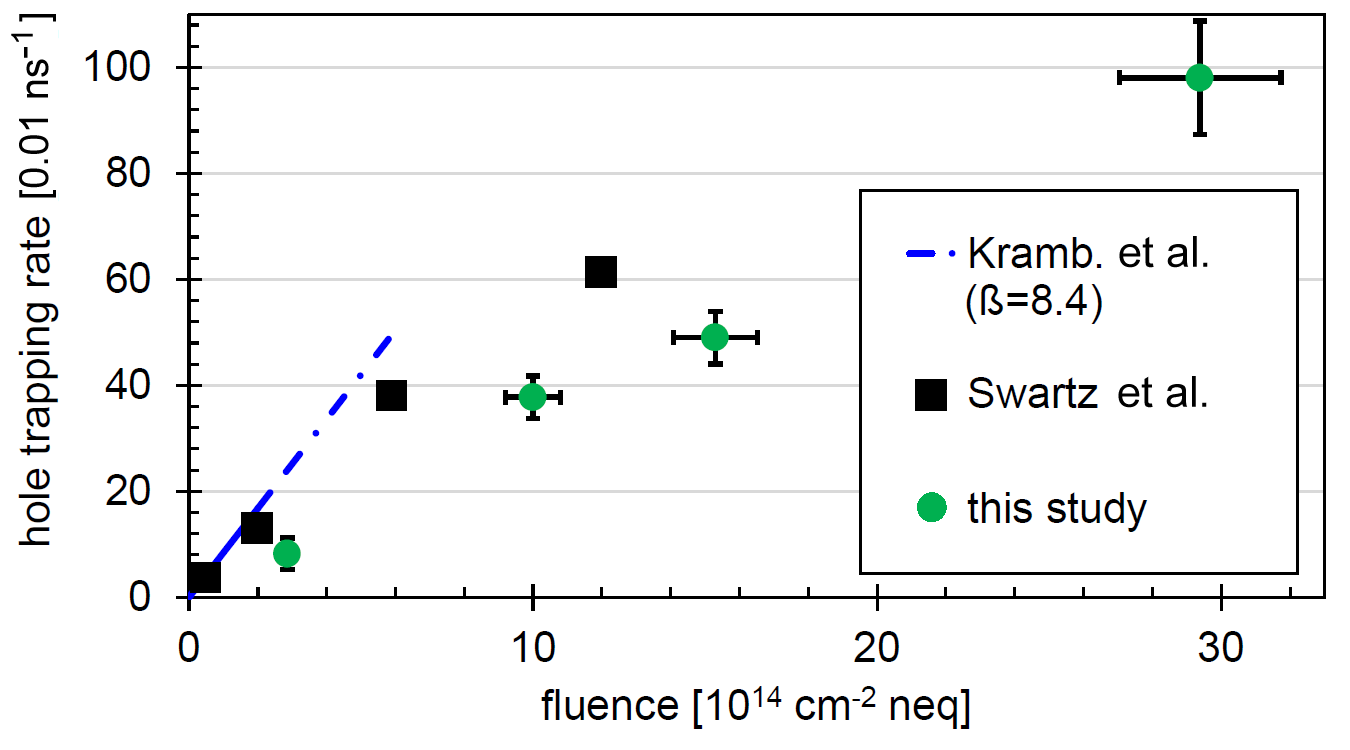}	
       \caption{
			Effective trapping rates for electrons (left) and holes (right) at $V = 600$~V according to this study (23~GeV protons, $T=-20$\,$^\circ$C) compared to studies by G.~Kramberger et al.~\cite{Kramberger:2002} (24~GeV protons, $T=-10$\,$^\circ$C),  
				M.~Swartz et al.~\cite{PixelAV} (24~GeV protons, $T=-10$\,$^\circ$C), 
				and R. Eber~\cite{Eber:2013} (23~MeV protons, $T=-10$\,$^\circ$C). The vertical error bars show the CCE uncertainties 
				(Table \ref{tab:trapping_rates}).  No significant difference has been observed between $T=-10$\,$^\circ$C and $T=-20$\,$^\circ$C.
			}
	\label{fig:trapping_rates}
\end{figure}

The effective trapping rates that provide the best description of the measurements are listed in Table~\ref{tab:trapping_rates}. 
They are also shown in Figure~\ref{fig:trapping_rates}, where they are compared to the trapping rates reported in Refs.~\cite{Kramberger:2002, Eber:2013, PixelAV}. For $3\cdot10^{14}$~neq/cm$^2$ the results show little dependence on the electric field: they are the same regardless of whether the parameters for $2\cdot10^{14}$~neq/cm$^2$ or $6\cdot10^{14}$~neq/cm$^2$ (Table~1) are used. For $3\cdot10^{15}$~neq/cm$^2$ the electric field was calculated according to Ref.~\cite{PixelAV} with the electric field tuned to describe pixel sensors irradiated with a similar fluence of $2.4\cdot10^{15}$~neq/cm$^2$. The CCE uncertainties quoted in Table~\ref{tab:trapping_rates} are the statistical and systematic uncertainties added in quadrature. To determine the E-field uncertainty different electric field distributions are tested, for each fluence the two that correspond to the closest fluences available, e.g. for $1\cdot10^{15}$~neq/cm$^2$ the values in Table~1 are used to calculate the electric fields that correspond to $6\cdot10^{14}$~neq/cm$^2$ and to $1.2\cdot10^{15}$~neq/cm$^2$. The CCE measurements taken at the highest fluence of $1.3\cdot10^{16}$~neq/cm$^2$ (Figure~1) are not analysed, since no simulation of the electric field at similar fluences is available.

For electron trapping the results of this work are compatible with the results presented in Refs.~\cite{Kramberger:2002, Eber:2013, PixelAV} for the fluences studied there. However, it is clear that the results presented in Ref.~\cite{Kramberger:2002} cannot be extrapolated to fluences of $10^{15}$~neq/cm$^2$ and above as this leads to an overestimate of trapping rates (and consequently an underestimate of the CCE). This has already been observed in Refs.~\cite{Lan09,Eber:2013}. For hole trapping lower trapping rates compared to Ref.~\cite{Kramberger:2002} are already observed at $3\cdot10^{14}$~neq/cm$^2$. This corresponds to a high CCE at this fluence (Figure~\ref{fig:CCE_red_front}). 

For one irradiation ($1\cdot10^{15}$~neq/cm$^2$ of 23~GeV protons) the electron trapping rate was extracted not only at 600~V but also at 400~V and at 900~V. For this fluence we do not expect charge multiplication below 1000~V, since charge multiplication starts to be relevant only above 120~kV/cm~\cite{Poe10}. Simulated fields are below 70 kV/cm at 600 V (Figure~\ref{fig:E-fields}). It is found that the trapping rates are similar, but slightly higher at 400~V (0.33~ns$^{-1}$) and slightly smaller at 900~V (0.28 ns$^{-1}$) compared to the rate at 600~V (0.3~ns$^{-1}$). Similar effects have been reported in Refs.~\cite{Lan09,Poe10}.

\begin{table}
\centering
\begin{tabular}{cc@{}c@{}@{}c@{}cc@{}c@{}c@{}}

$\phi_{neq}$ $[$neq/cm$^2]$ 			& \multicolumn{3}{c}{ $1/\tau_e$~$[1/ns]$	} 
																	&& \multicolumn{3}{c}{ $1/\tau_h$~$[1/ns]$	} \\  \hline  
$3	\cdot 10^{14}$	 	&	$0.145		$&\,$\pm 0.035{(\text{CCE})} 	$&\,$\pm 0.005{(\text{E-field})}   $		
											&&	$0.085		$&\,$\pm 0.025{(\text{CCE})} 	$&\,$\pm 0.005{(\text{E-field})}   $ \\ 
$1	\cdot 10^{15}	$ 	&	$0.30 	$&\,$\pm 0.04{(\text{CCE})} 	$&\,$\pm 0.03{(\text{E-field})} $	
											&&	$0.38			$&\,$\pm 0.04{(\text{CCE})} 	$&\,$\pm 0.04{(\text{E-field})}   $ \\ 

$1.5	\cdot 10^{15}$	& $0.42		$&\,$\pm 0.04{(\text{CCE})} 	$&\,$\pm 0.03{(\text{E-field})} $	 
											&& $0.49		$&\,$\pm 0.05{(\text{CCE})} 	$&\,$\pm 0.03{(\text{E-field})} $	\\

$3	\cdot 10^{15}$	 	&	$0.55		$&\,$\pm 0.06{(\text{CCE})} 	$&\,$\pm 0.06{(\text{E-field})} $ 
											&&	$0.98		$&\,$\pm 0.10{(\text{CCE})} 	$&\,$\pm 0.12 {(\text{E-field})} $ \\
 \end{tabular}
   \caption{Trapping rates extracted from the CCE at $T=-20^\circ$\,C and $V=600$~V after irradiation with 23~GeV protons. Note that the trapping rates are effective rates describing the CCE for electrons drifting from the p-n junction to the rear side, and holes drifting from the rear side to the p-n junction. 
	}
   \label{tab:trapping_rates}
\end{table}

\section{Applicability for eh-pairs generated along the sensor depth}
To test the applicability of the results reported in Table~\ref{tab:trapping_rates} for cases where $eh$-pairs are generated along the whole sensor depth (as is the case for charged particles traversing the sensor) further simulations are performed and compared to the measurements. 

\begin{figure}
	\centering
	\includegraphics[width=8cm]{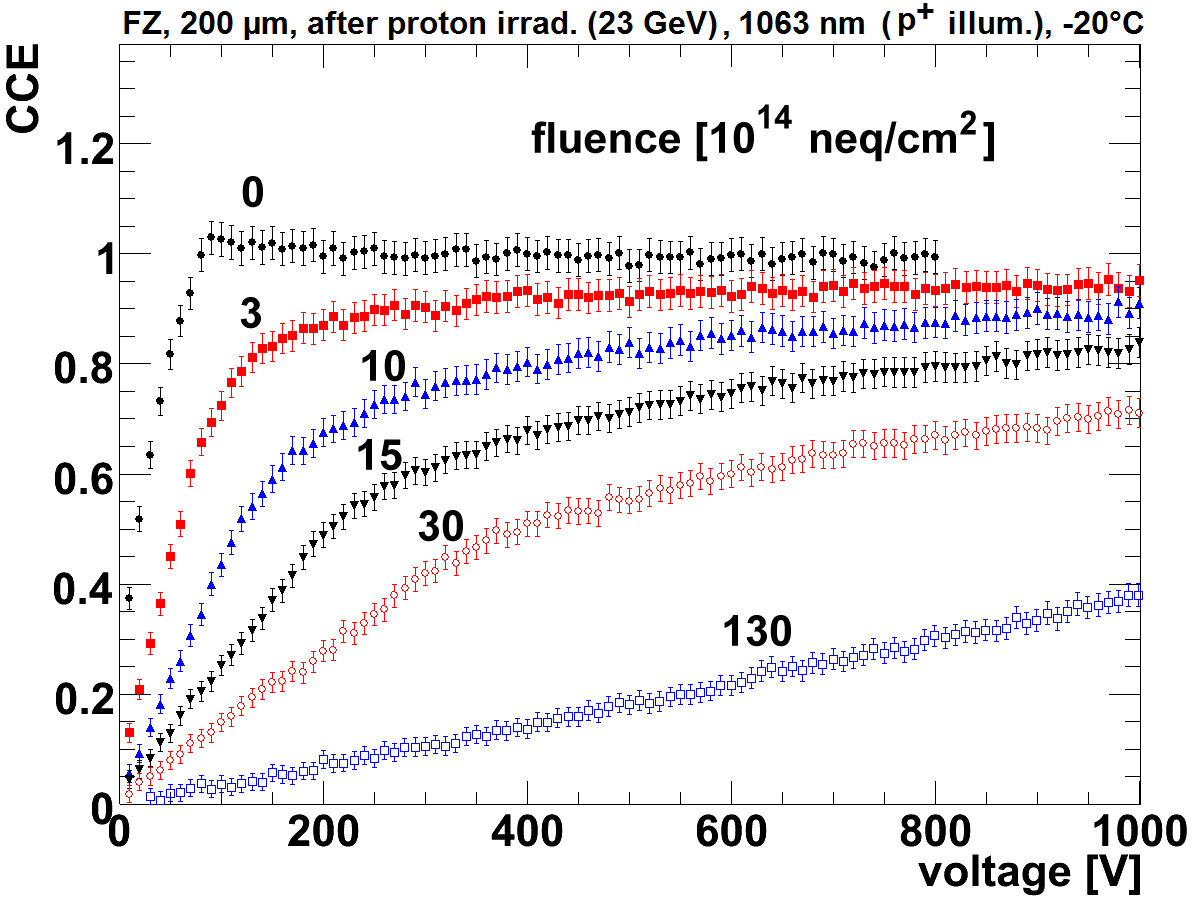}
  \caption{The CCE as a function of bias voltage for 200~$\mu$m thick n-type sensors after irradiation with different fluences 
	of 23~GeV protons. 
	Laser light of 1062~nm wavelength was used to generate $eh$-pairs throughout the whole sensor depth.}
	\label{fig:CCE_IR}
\end{figure}

A separate simulation is performed in which $eh$-pairs are generated along the whole sensor depth using an attenuation length of 1\,000~$\mu$m. The simulation was used to describe CCE measurements where $eh$-pairs are generated using light of 1062~nm wavelength (front-side illumination). The measured CCE as a function of bias voltage is shown in Figure~\ref{fig:CCE_IR} and a comparison of the simulated and measured CCE at 600~V bias is presented in Table~\ref{tab:CCE_IR}.

The simulated CCE is, on average, 0.06 below the measured CCE if the trapping rates from Table~\ref{tab:trapping_rates} (simulation~A) are used. This indicates that the effective trapping rates for the measurements using 1062~nm light are lower than for the measurements where $eh$-pairs are generated close to the implants only. This is expected if the leakage current leads to a non-uniform occupation of defects that are relevant for trapping.
We conclude that charge losses might be overestimated if the rates given in Table~\ref{tab:trapping_rates} are used to predict charge collection in cases where $eh$-pairs are generated along the whole sensor depth. However, compared to the widely used extrapolation of effective trapping rates at low fluences the overestimation is significantly reduced (Table~\ref{tab:CCE_IR}).

The results presented here may be used in further simulations of irradiated silicon sensors. Due to the complex (and sometimes non-linear) generation of defects in the irradiation process we have not described the results with a parameterisation. 
Instead a linear interpolation may be used to simulate fluences lying in between those considered in this work. For significantly different bias voltages a correction may be applied (effective trapping rates are about 10~\% higher when the bias voltage is decreased by 200V). 
The applicability of the results was not tested using segmented sensors. However, comparison with other studies~\cite{PixelAV} shows that similar values of the effective trapping rates are used to describe data in strip sensors (Figure~\ref{fig:trapping_rates}). 

\begin{table}
\centering
\begin{tabular}{cccc}

$\phi_{neq}$ $[$neq/cm$^2]$ 	& CCE measured  			& CCE simulated A & CCE simulated B \\  \hline
$3	\cdot 10^{14}$	 					& $0.93 \pm 0.03$			&	$0.92 \pm 0.03$	& $0.85 \pm 0.02$	\\
$1	\cdot 10^{15}$	 					& $0.85 \pm 0.03$			&	$0.76 \pm 0.03$	& $0.60 \pm 0.03$	\\
$1.5	\cdot 10^{15}$	 				& $0.75 \pm 0.03$			&	$0.70 \pm 0.03$	& $0.50 \pm 0.03$	\\
$3	\cdot 10^{15}$	 					& $0.60 \pm 0.03$			&	$0.51 \pm 0.03$	& $0.27 \pm 0.02$	\\
 \end{tabular}
   \caption{Simulated values of the CCE compared to measurements at 600~V using 1062~nm light to generate $eh$-pairs. In simulation~A, the trapping parameters from Table~\ref{tab:trapping_rates} are used, in simulation~B, extrapolated trapping rates (Equation~\ref{eq:kramberger}, $\beta_{e} = 5.8 \cdot10^{-16}$cm$^2$/ns, $\beta_{h} = 8.2 \cdot10^{-16}$cm$^2$/ns) according to Ref.~\cite{Kramberger:2002} are used.} 
   \label{tab:CCE_IR}
\end{table}

\section{Summary}

Time-resolved charge collection measurements using red laser light of 672~$\mu$m wavelength have been used to determine the effective trapping rates for electrons (holes) moving from the p$^+$ to the n$^+$ (n$^+$ to the p$^+$) contact in silicon single-pad sensors irradiated with protons with fluences up to $3\cdot10^{15}$~neq/cm$^2$. Light of this wavelength has a penetration depth of about 3.5~$\mu$m in silicon. 
The time-resolved measurements are described using simulation. The electric fields have been calculated assuming two effective traps with energy levels, concentrations, and cross-sections taken from the literature \cite{Eber:2013, EVL2, PixelAV}. 

It is found that at the lowest investigated fluence ($3\cdot10^{14}$~neq/cm$^2$) the effective electron trapping rate is compatible with the results presented in Ref.~\cite{Kramberger:2002} using fluences up to $2.4\cdot10^{14}$~neq/cm$^2$. However, at higher fluences the extracted trapping rates are a factor of 2--3 below the trapping rates expected if the results from Ref.~\cite{Kramberger:2002} are extrapolated. The effective hole trapping rates are also a factor of up to about 3 below the extrapolations.
These results confirm previous studies that found higher signals than expected at high fluences \cite{Lan09,Poe10,Eber:2013} and are important for the description of the CMS Tracker performance after a few years of operation at the High-Luminosity LHC.

\begin{acknowledgments}
The research leading to these results has received funding from the European Commission under the FP7 Research Infrastructures project AIDA, grant agreement no. 262025. The information herein only reflects the views of its authors and not those of the European Commission and no warranty expressed or implied is made with regard to such information or its use.
Support was also provided by the Helmholtz Alliance ``Physics at the Terascale'' and the German Ministry of Science, BMBF, through the Forschungsschwerpunkt ``Particle Physics with the CMS-Experiment''. 
\end{acknowledgments}

\bibliography{auto_generated}   

\begin{thebibliography}{10}
\bibitem{MS}
\hrefCMSnoop {} {M.~Swartz}. Private communication. The data for the fluence
  point $12\cdot10^{14}$~neq/cm$^2$ has not been published. It was determined
  using the same methods as described in Ref.~\cite{PixelAV}.
\end{thebibliography}

\newpage
\appendix
\section{\cmsCollabName \label{app:collab}}

\begin{sloppypar}\hyphenpenalty=5000\widowpenalty=500\clubpenalty=5000\flushleft
\textbf{Institut f\"ur Hochenergiephysik der \"Osterreichischen Akademie der Wissenschaften (HEPHY), Vienna, Austria}\\*[0pt]
W.~Adam, T.~Bergauer, M.~Dragicevic, M.~Friedl, R.~Fruehwirth, M.~Hoch, J.~Hrubec, M.~Krammer, W.~Treberspurg, W.~Waltenberger
% ...
\end{sloppypar}
\textbf{Corresponding Author:} Thomas Poehlsen, e-mail: thomas.poehlsen@cern.ch
\end{document}